\documentclass[final,authoryear,3p,times,10pt]{elsarticle}

\usepackage{multirow,setspace,times,amssymb,amsmath,graphicx,color,rotating,subfigure,url}
\usepackage{lineno}
\usepackage{natbib}
\usepackage{mathrsfs}
\usepackage{makecell}
\usepackage{CJK}
\usepackage[CJKbookmarks=true,colorlinks,linkcolor=red,anchorcolor=blue,citecolor=green,unicode]{hyperref} %
\usepackage{bookmark}

\journal{Elsevier}

\begin{document}

\begin{frontmatter}

\title{Time series classification based on triadic time series motifs}

\author[SB,RCE]{Wen-Jie Xie}
\author[DM]{Rui-Qi Han}
\author[SB,RCE,DM]{Wei-Xing Zhou\corref{WXZ}}
\cortext[WXZ]{Corresponding to: 130 Meilong Road, P.O. Box 114, School of Business, East China University of Science and Technology, Shanghai 200237, China.}
\ead{wxzhou@ecust.edu.cn}

\address[SB]{Department of Finance, East China University of Science and Technology, Shanghai 200237, China}
\address[RCE]{Research Center for Econophysics, East China University of Science and Technology, Shanghai 200237, China}
\address[DM]{Department of Mathematics, East China University of Science and Technology, Shanghai 200237, China}

\begin{abstract}
It is of great significance to identify the characteristics of time series to qualify their similarity. We define six types of triadic time-series motifs and investigate the motif occurrence profiles extracted from logistic map, chaotic logistic map, chaotic Henon map, chaotic Ikeda map, hyperchaotic generalized Henon map and hyperchaotic folded-tower map. Based on the similarity of motif profiles, we further propose to estimate the similarity coefficients between different time series and classify these time series with high accuracy. We further apply the motif analysis method to the UCR Time Series Classification Archive and provide evidence of good classification ability for some data sets. Our analysis shows that the proposed triadic time series motif analysis performs better than the classic dynamic time wrapping method in classifying time series for certain data sets investigated in this work.
\end{abstract}


\end{frontmatter}

\section{Introduction}
\label{S1:Introduction}

Quantifying the similarity of time series has always been a very useful primitives for time series analysis, with applications to many fields \citep{Hu-Chen-Keogh-2016-DMKD,Silva-Souza-Ellis-Keogh-Batista-2015-JIRS,Gomes-Batista-2015-ASTL,Mueen-Keogh-Zhu-Cash-Westover-2009-SIAM,Mcgovern-Rosendahl-Brown-Droegemeier-2011-DMKD,Chiu-Keogh-Lonardi-ACM,Mueen-Keogh-2010-KDD,Tataw-Reddy-Keogh-Roy-Chowdhury-2013-TCBB}.
The key point of measuring similarity is to define a suitable and effective distance between two time series \citep{Hu-Rakthanmanon-Campana-Mueen-Keogh-2015-PAA,Tarango-Keogh-Brisk-2014-ACSSC,Miskiewicz-Ausloos-2008-PA}.
The widely adopted definitions of distance include the Euclidean distance and correlation measures \citep{Miskiewicz-Ausloos-2008-PA}.
However, in terms of measuring the similarity of time series, the Euclidean distance is often average, sometimes bad \citep{Wang-Mueen-Ding-Trajcevski-Scheuermann-Keogh-2013}.
For most time series analysis problems, the dynamic time warping (DTW) provides a highly competitive distance metric \citep{Silva-Giusti-Keogh-Batista-2018-DMKD,Wang-Mueen-Ding-Trajcevski-Scheuermann-Keogh-2013}.
To get the best performance of DTW, we need to regulate its unique parameter to optimize the dynamic time warping's window width \citep{Dau-Silva-Petitjean-Forestier-Bagnall-Mueen-Keogh-2018-DMKD}.
The complexity of the DTW method is relatively high, so many researchers provide some improved methods to have better performance \citep{Petitjean-Forestier-Webb-Nicholson-Chen-Keogh-2016-KIS,Dau-Silva-Petitjean-Forestier-Bagnall-Keogh-2017-IEEE,Mueen-Keogh-2016-SIGKDD}.
Moreover, practitioners generalize the DTW to some multi-dimensional time series classification experiments \citep{Shokoohi-Yekta-Hu-Jin-Wang-Keogh-2017-DMKD}.

Similar subsequences in time series can be defined as time series motifs, which characterize the temporal properties and dynamics of the corresponding long time series
\citep{Mcgovern-Rosendahl-Brown-Droegemeier-2011-DMKD,Chiu-Keogh-Lonardi-ACM,Mueen-Keogh-2010-KDD}.
It is useful for exploratory data mining and often used as inputs for classification of time series, clustering, segmentation
\citep{Bagnall-Lines-Bostrom-Large-Keogh-2017-DMKD,Mori-Mendiburu-Keogh-Lozano-2017-DMKD,Dau-Begum-Keogh-2016-CIKM,Petitjean-Forestier-Webb-Nicholson-Chen-Keogh-2014-ICDM}.
Time series motif analysis has been widely used in diverse fields
\citep{Yeh-Zhu-Ulanova-Begum-Ding-Dau-Zimmerman-Silva-Mueen-Keogh-2018-DMKD,Zhu-Zimmerman-Senobari-Yeh-Funning-Mueen-Brisk-Keogh-2018-KIS,Linardi-Zhu-Palpanas-Keogh-2018-SIGMOD,Linardi-Zhu-Palpanas-Keogh-2018-SIGMODb,Yeh-Kavantzas-Keogh-2017-IEEE,Zakaria-Mueen-Keogh-Young-2016-DMKD}.
Gomes and Batista presented a SAX-based motif discovery method to classify the urban sound \citep{Gomes-Batista-2015-ASTL}.
Wang et al. proposed a method to automatically detect repeating segments in music and two time series data sets \citep{Wang-Chng-Li-2010-PRL}.
Son and Anh introduced two novel methods to discover approximate $k$-motifs in time series data \citep{Son-Anh-2016-KIS}
and their methods play an important role in several time series data mining tasks by using motif discovery.
Lots of researchers have used time series motifs analysis for applications in many different domains \citep{Mueen-Keogh-Zhu-Cash-Westover-2009-SIAM,Mcgovern-Rosendahl-Brown-Droegemeier-2011-DMKD,Chiu-Keogh-Lonardi-ACM,Mueen-Keogh-2010-KDD}.

Triadic time series motifs \citep{Xie-Han-Zhou-2019-EPL} are inspired by the network motifs in visibility graph \citep{Lacasa-Luque-Luque-Nuno-2009-EPL,Lacasa-Luque-Ballesteros-Luque-Nuno-2008-PNAS,Ni-Jiang-Zhou-2009-PLA,Yang-Wang-Yang-Mang-2009-PA,Elsner-Jagger-Fogarty-2009-GRL,Qian-Jiang-Zhou-2010-JPA} and horizontal visibility graphs (HVG) mapping from time series \citep{Lacasa-Luque-Luque-Nuno-2009-EPL,Elsner-Jagger-Fogarty-2009-GRL,Lacasa-Toral-2010-PRE,Shao-2010-APL,Dong-Li-2010-APL,Ahmadlou-Adeli-Adeli-2010-JNT,Tang-Liu-Liu-2010-MPLB,Xie-Han-Jiang-Wei-Zhou-2017-EPL,Xie-Han-Zhou-2019-CNSNS}.
The six triadic time series motifs are similar in some features with sequential HVG motifs \citep{Iacovacci-Lacasa-2016a-PRE,Iacovacci-Lacasa-2016b-PRE} and ordinal patterns \citep{Keller-Sinn-2005-PA,McCullough-Small-Stemler-Iu-2015-Chaos, McCullough-Small-Iu-Stemler-2017-PTA, Zhang-Zhou-Tang-Guo-Small-Zou}. The permutation entropy based on ordinal patterns \citep{Bandt-Pompe-2002-PRL,Amigo-2010} is a natural complexity measure and useful in the presence of dynamical or observational noise.
Similarly, the triadic time series motif analysis can also mine the dynamical characteristics of time series from complex system.
\cite{Xie-Han-Zhou-2019-EPL} used the triadic time series motif analysis to uncover the different dynamics in the heartbeat rates of healthy subjects, congestive heart failure subjects, and atrial fibrillation subjects and identify the bullish and bearish markets from the price fluctuations of financial markets.

In this work, we identify six triadic time series motifs and investigate their occurrence profiles in time series from logistic maps with different control paratemers and chaotic time series generated from chaotic logistic map, chaotic Henon map, chaotic Ikeda map, hyperchaotic generalized Henon map and hyperchaotic folded-tower map.
It is of great significance to be able to discover the characteristics of time series from different types of chaotic maps.
We also apply the triadic time series motif analysis to classify the time series in 128 data sets from UCR Time Series Classification Archive \citep{UCRArchive2018}.

\section{Triadic time series motifs}
\label{S1:Methods}

\begin{figure}[ht]
  \centering
  \includegraphics[width=7cm]{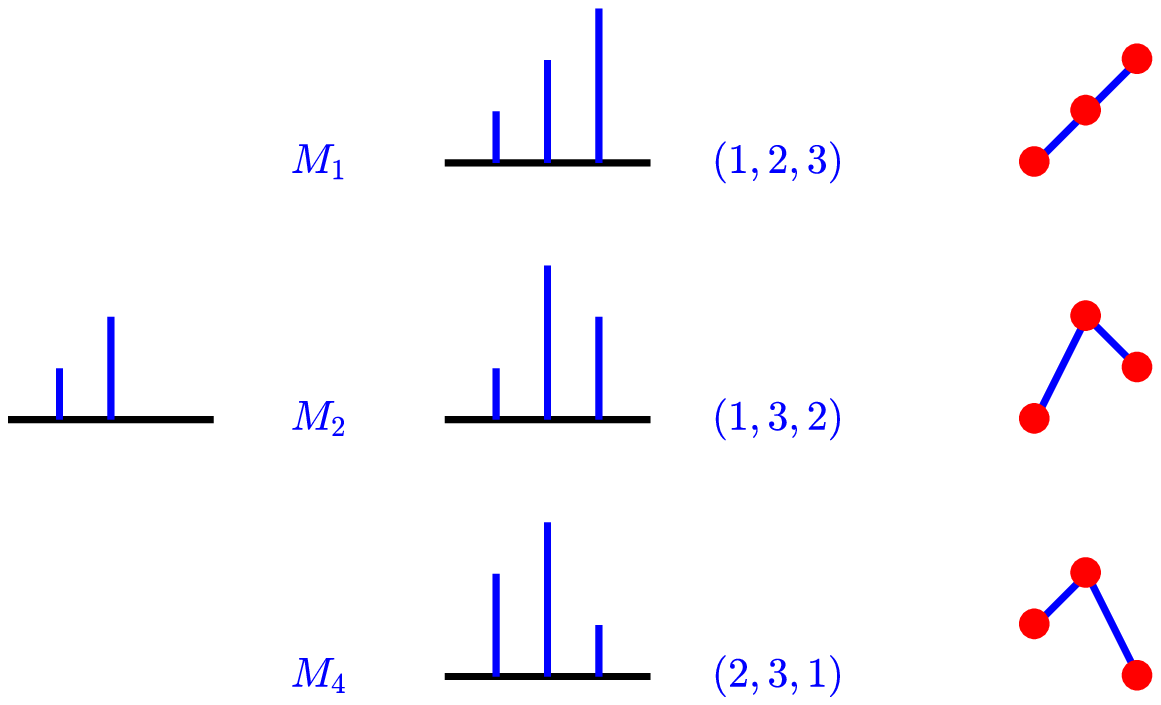}
  \hskip 1.2cm
  \includegraphics[width=7cm]{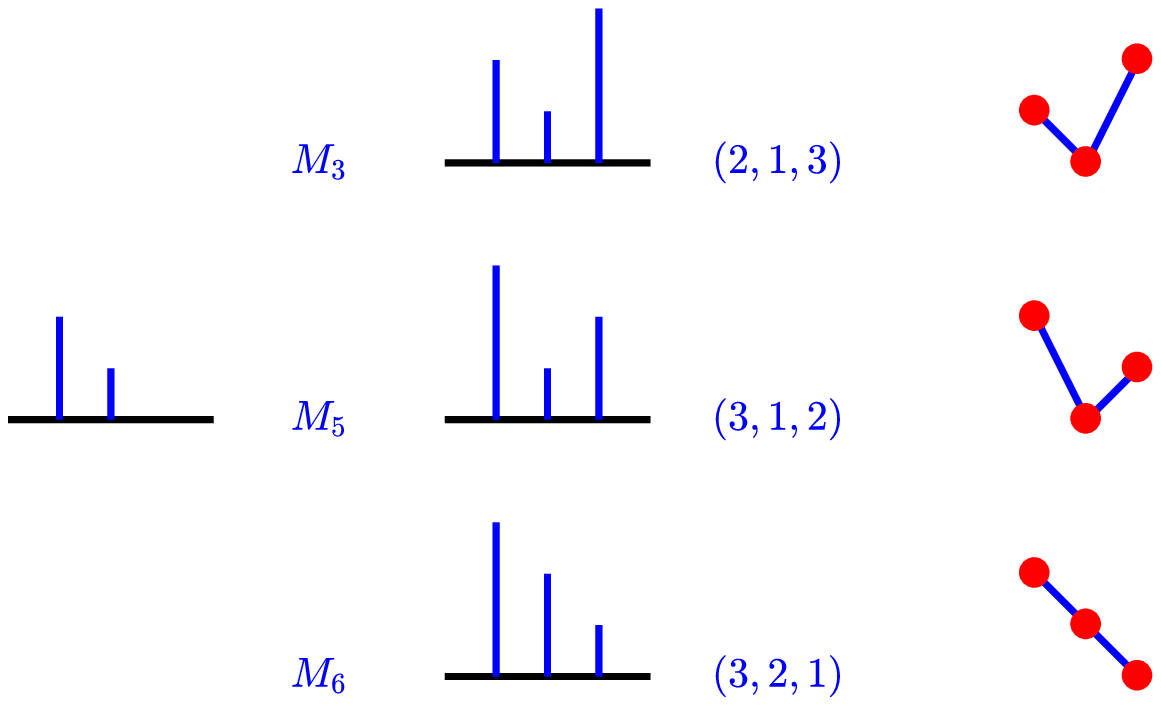}
  \caption{\label{Fig:TSMotif:3N:Rand:Motif:Plot:Defi} Illustrative example showing the six types of triadic motifs in time series.}
\end{figure}

Triadic time series motifs are determined by the relative magnitude and ordinal order of three data points that are randomly chosen from the time series \cite{Xie-Han-Zhou-2019-EPL}. For three arbitrary data $\{x_{i}, x_{j}, x_{k}\}$ with $i<j<k$ in the time series $\{x_{i}\}_{i = 1,...,L}$, a time series motif forms if the following conditions is fulfilled \citep{Xie-Han-Zhou-2019-EPL}:
\begin{equation}
   \left\{
    \begin{array}{lllll}
     x_i>x_n &{\rm{and}}& x_j>x_n, &  \forall n\in(i,j) \\
     x_j>x_m &{\rm{and}}& x_k>x_m, &  \forall m\in(j,k)
    \end{array}
    \right..
    \label{Eq:TSMotif}
\end{equation}
We obtain six triadic time series motifs, which are denoted as $M_1$, $M_2$, $M_3$, $M_4$, $M_5$, $M_6$ in Fig.~\ref{Fig:TSMotif:3N:Rand:Motif:Plot:Defi}. This definition does not consider situations where two or three data points of $\{x_{i}, x_{j}, x_{k}\}$ are equal. When two data points are identical, we treat it as if the latter data point is larger than the former one.

The time series motifs are different from the conventional motifs of horizontal visibility graphs \citep{Lacasa-Toral-2010-PRE,Lacasa-Luque-Luque-Nuno-2009-EPL,Shao-2010-APL,Dong-Li-2010-APL,Ahmadlou-Adeli-Adeli-2010-JNT,Elsner-Jagger-Fogarty-2009-GRL,Tang-Liu-Liu-2010-MPLB,Xie-Zhou-2011-PA,Xie-Han-Jiang-Wei-Zhou-2017-EPL,Xie-Han-Zhou-2019-CNSNS}. Considering the triadic HVG motif, there are only two admissible motifs in undirected HVGs, one being a chain and the other being a triangle. As shown in Fig.~\ref{Fig:TSMotif:3N:Rand:Motif:Plot:Defi}, the open triadic motif can be mapped from the time series $(1,2,3)$, $(1,3,2)$, $(2,3,1)$ and $(3,2,1)$ and the close triadic motif can be mapped from the time series $(2,1,3)$ and $(3,1,2)$. Time series motifs consider not only the visibility between data points, as HVG motifs, but also the order and relative magnitudes of the points. Hence, time series motifs explore finer structures of HVG motifs \citep{Xie-Han-Zhou-2019-EPL}.

\section{Triadic time series motif analysis of chaotic maps}

\subsection{Chaotic maps}

We perform triadic time series motif analysis numerically for different time series in continuous and discrete dynamic systems. Through extensive numerical experiments, we investigate the motif distribution extracted from the logistic map, the chaotic logistic map, the chaotic Henon map, the chaotic Ikeda map, the hyperchaotic generalized Henon map, and the hyperchaotic folded-tower map.

The logistic map is a representative example of how complex, chaotic behaviour can arise from very simple nonlinear dynamical equation. Mathematically, the logistic map is written as
\begin{equation}
x_{n+1}=rx_{n}(1-x_{n}).
\end{equation}

To distinguish chaotic maps and hyperchaotic maps, we generate four types of time series from chaotic Henon map, chaotic Ikeda map, hyperchaotic generalized Henon map and hyperchaotic folded-tower map.
The specific equations for these four types of dynamic systems are given below \citep{Xu-Zhang-Small-2008-PNAS}.
Mathematically, the chaotic Henon map is written as
\begin{equation}
     \left\{
    \begin{aligned}
      x_{n+1}&=y_n+1-ax_n^2,\\
      y_{n+1}&=bx_n.
    \end{aligned}
    \right.
\end{equation}
where $a=1.4$ and $b=0.3$.
The chaotic Ikeda map is written as
\begin{equation}
     \left\{
    \begin{aligned}
      x_{n+1}&=1+\mu(x_n \cos t_n-y_n \sin t_n),\\
      y_{n+1}&=\mu(x_n \sin t_n+y_n \cos t_n).
    \end{aligned}
    \right.
\end{equation}
where $t_n=0.4-6/(1+x_n^2+y_n^2)$ and $\mu=0.7$.
The hyperchaotic generalized Henon map is written as
\begin{equation}
     \left\{
    \begin{aligned}
      x_{n+1}&=a- y_n^2-bz_n,\\
      y_{n+1}&=x_n,\\
      z_{n+1}&=y_n.
    \end{aligned}
    \right.
\end{equation}
where $a=1.9$ and $b=0.03$.
The hyperchaotic folded-tower map is written as
\begin{equation}
     \left\{
    \begin{aligned}
      x_{n+1}&=ax_n(1-x_n)- 0.05(y_n+0.35)(1-2z_n),\\
      y_{n+1}&=0.1((y_n+0.35)(1+2z_n)-1)(1-1.9x_n),\\
      z_{n+1}&=3.78z_n(1-z_n)+by_n.
    \end{aligned}
    \right.
\end{equation}
where $a=3.8$ and $b=0.2$.

\subsection{Occurrence frequency distributions of triadic motifs}

\begin{figure}[t!]
\centering
  \includegraphics[width=15cm]{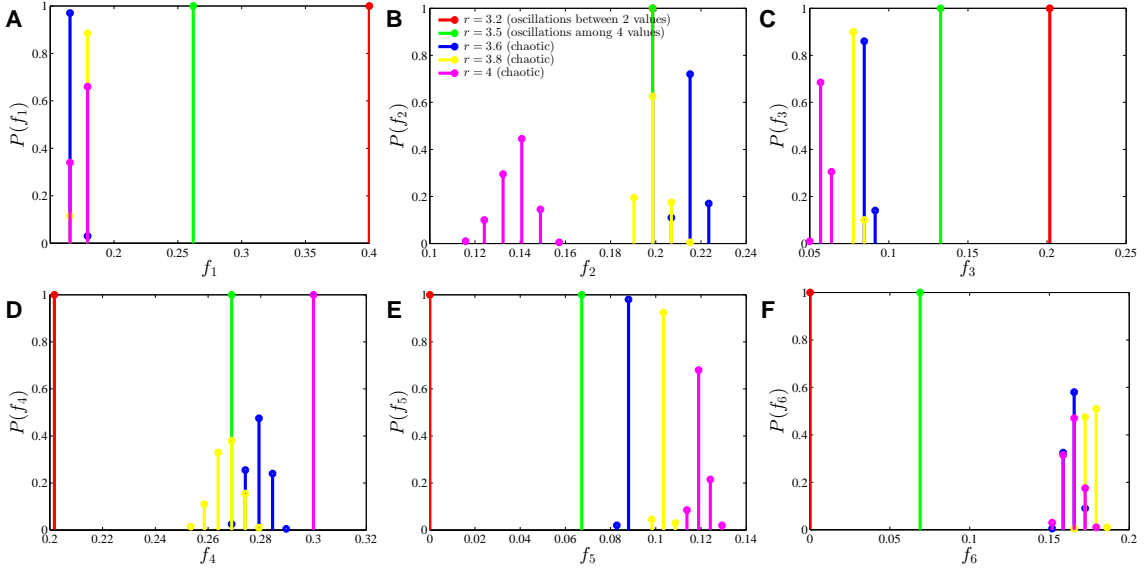}
  \caption{The probability distribution of occurrence frequency $f_i$ of motifs $M_i$ for different types of time series of the logistic map with parameter $r=3.2, ~3.5, ~3.6, ~3.8, ~4$. }
    \label{Fig:TSMotif:Mimic:PDFlogistic}
\end{figure}

We first generate time series by using the logistic map with control parameter $r$. The parameter $r$ ranges in the interval of $(0, 4]$.
When $r=3.2$ , $x$ will approach permanent oscillations between two values from almost all initial conditions.
When $r=3.5$, $x$ will approach permanent oscillations among four values from almost all initial conditions.
When $r=3.6$, $3.8$, or $ 4$, $x$ exhibits chaotic behaviour.
For each parameter $r$, we generate 200 time series with length $512$, determine the occurrence frequencies $f_i$ of the six triadic time series motifs, and obtain the occurrence frequency distribution of each motif. Fig.~\ref{Fig:TSMotif:Mimic:PDFlogistic} shows the distributions of the occurrence frequency $f_i$ of the motif $i$ in the logistic time series with $r=3.2$, $3.5$, $3.6$, $3.8$, and $4$. We find that the five classes of time series have very different occurrence frequency distributions of time series motifs.

When the parameter $r=3.2$, the time series is $\{x_{i}\}_{i = 1,...,L}=\{a, b, a, b, a, b,\cdots\}$. Without loss of generality, we assume that $a>b$. The set of motif $M_1$ is the union of $\{x_{2i-1}, x_{2i+1}, x_{2i+3}\}$ and $\{x_{2i-2}, x_{2i}, x_{2i+2}\}$ with $i=1,2,...,[L/2]-2$, so that the occurrence count of $M_1$ is
\begin{equation}
  O(M_1) = 2[L/2]-4.
\end{equation}
The sets of motifs $M_2$, $M_3$ and $M_4$ are respectively the union of $\{x_{2i}, x_{2i+1}, x_{2i+2}\}$, of $\{x_{2i-1}, x_{2i+1}, x_{2i+2}\}$, and of $\{x_{2i-1}, x_{2i}, x_{2i+1}\}$, where $i=1,2,...,[L/2]-1$. It follows that
\begin{equation}
  O(M_2) = O(M_3) = O(M_4) = 2[L/2]-2.
\end{equation}
By definition, motifs $M_5$ and $M_6$ cannot appear. Therefore, the occurrence frequencies
\begin{equation}
  f_i := f(M_i)= \lim_{L\to\infty}\frac{O(M_i)}{\sum_{i=1}^6O(M_i)}
\end{equation}
are obtained as follows
\begin{equation}
  {\mathbf{f}} := [f_1, \cdots, f_6] = [0.4, 0.2, 0.2, 0.2, 0, 0].
\end{equation}
This analytical result is verified by the numerical simulations, as shown in Fig.~\ref{Fig:TSMotif:Mimic:PDFlogistic} (red bars).

When the parameter $r=3.5$, the time series is slightly more complicated. As shown in Fig.~\ref{Fig:TSMotif:Mimic:PDFlogistic}, for the same motif, the distribution of occurrence frequency $f_i$ of the motif $M_i$ is concentrated and the variance is small, even 0.
When $r>3.54$, the oscillation period becomes longer and longer, until about $r=3.6$, the period tends to infinity, and the system becomes a chaotic system. When $r>3.6$, the result of the iterative run will switch between the period type and the chaotic type. Until $r=4$, the system is complete chaos. Although they are all chaos, the distribution of occurrence frequency $f_i$ for $r=3.6, ~3.8, ~4$ in Fig.~\ref{Fig:TSMotif:Mimic:PDFlogistic} has a big difference.

\begin{figure}[h!]
\centering
\includegraphics[width=15cm]{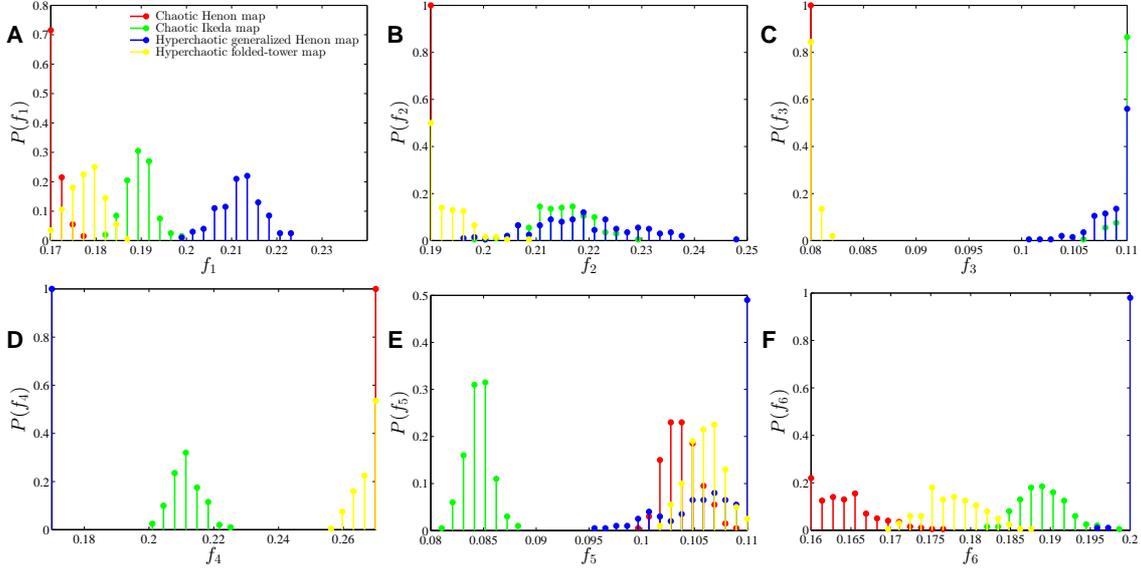}
\caption{The probability distribution of occurrence frequency $f_i$ of motifs $M_i$ for different types of time series: chaotic Henon map, chaotic Ikeda map, hyperchaotic generalized Henon map and hyperchaotic folded-tower map . The length of time series is 512.}
\label{Fig:TSMotif:Mimic:ChaoticPDF}
\end{figure}

We further perform triadic time series motif analysis of the four types of discrete chaotic time series: chaotic Henon map, chaotic Ikeda map, hyperchaotic generalized Henon map and hyperchaotic folded-tower map. In Fig.~\ref{Fig:TSMotif:Mimic:ChaoticPDF}, the length of time series is 512 and there are big difference in the occurrence frequency of motif $M_3$ and motif $M_4$ between the four types of discrete chaotic time series.
It can be imagined that the longer the time series is, the larger the difference of the occurrence frequency of the individual motifs will be, and the easier it is to distinguish the the four types of discrete chaotic time series.
In Fig.~\ref{Fig:TSMotif:Mimic:ChaoticPDF}, we cannot use one motif's occurrence frequency to distinguish different types of chaotic time series, but in Fig.~\ref{Fig:TSMotif:Mimic:PDFlogistic}, we can use a single indicator $f_3$ to distinguish the five types of logistic time series. This method can be understood as a dimension reduction method, which reduces the time series with length $n$ to 5-dimensional space for different time series, because $\sum_{i=1}^6 f_i=1$.

\subsection{Classification of time series}

The triadic motif analysis is applied to the classification of time series to investigate the effectiveness of the similarity measure of time series.
In order to classify different time series, we need to extract the features of time series. The triadic time series motifs are used as the features of time series, and then the time series are classified based on the motif occurrence frequency distributions. From a common sense, the longer the time series, the more information obtained by the method for extracting features of time series, the more accurately the time series can be classified. Therefore, we consider the influence of time series lengths on the accuracy of classification. We compare two classical methods for measuring the similarity of time series: one is the simple Euclidean distance method, and the other is the dynamic time warping method (DTW). We select the nearest neighbor method (1NN) to classify the
time series based on the three similarity measures.

\begin{figure}[t!]
\centering
\includegraphics[width=14cm]{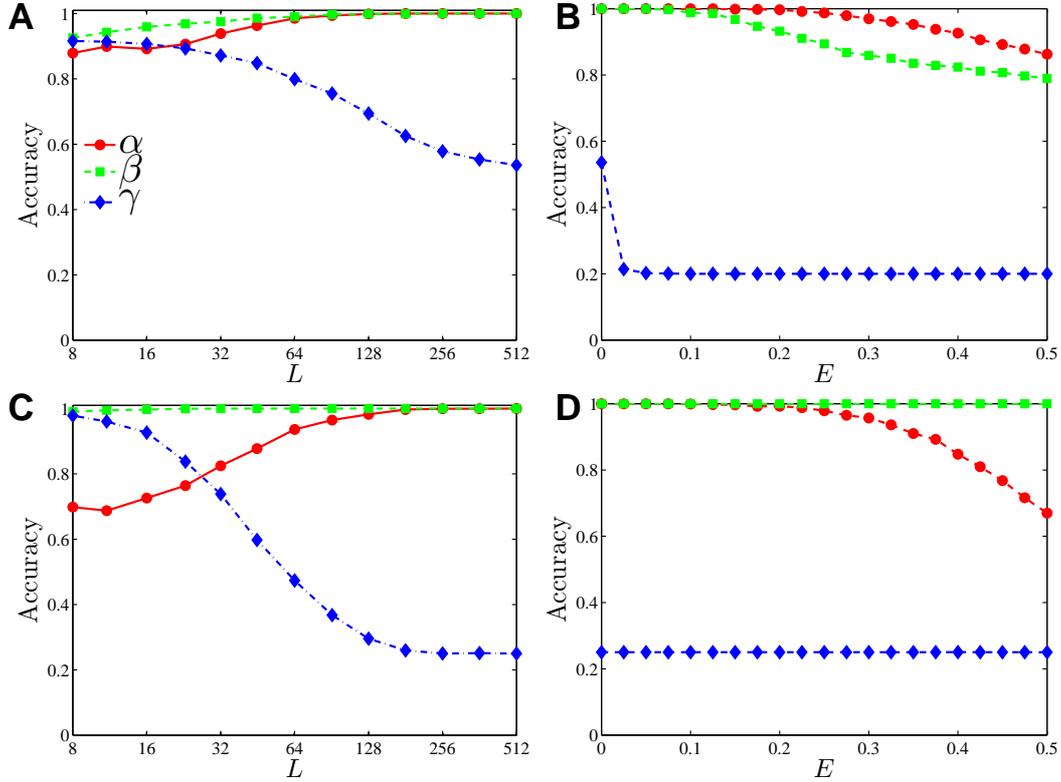}
\caption{The results of classification based on similarity of occurrence frequency of motifs for different types of time series.
(A) The average classification accuracy rate $\alpha$, $\beta$ and $\gamma$ for the five types of logistic time series with parameter $r=3.2$, $3.5$, $3.6$, $3.8$ and $4$. The time series contains 512 data points. For each type of logistic time series, we generates 100 time series training sets and 100 test sets.
The ordinate represents the average classification accuracy rate. The abscissa represents the length of time series.
The three colors (red, green, blue) correspond to three similarity measurements: the motif distribution, the DTW and Euclidean distance.
(B) The relationship between the data deletion rate $E$ and the average classification accuracy rate for the five types of logistic time series.
(C) The average classification accuracy rate  $\alpha$, $\beta$ and $\gamma$ for the chaotic Henon map, chaotic Ikeda map, hyperchaotic generalized Henon map and hyperchaotic folded-tower map, respectively.
(D) The relationship between the data deletion rate $E$ and the average classification accuracy rate for the four chaotic maps.
}
\label{Fig:TSMotif:Mimic:Classification}
\end{figure}

We analyze respectively the time series generated from the logistic map with parameter $r=3.2$, $3.5$, $3.6$, $3.8$ and $4$ and from the four chaotic maps. For each type of time series, we generate 100 time series as the training set and 100 time series as the test set.
To analyze the accuracy of classification of time series with different lengths $L$, the length of time series is changed from $L=8$ to $L=512$, and then the data sets are classified by the nearest neighbor method based on three similarity measures.
The three colors (red, green, blue) in Fig.~\ref{Fig:TSMotif:Mimic:Classification} (A) and (C) correspond to the three similarity measures: the motif occurrence profile, the DTW and the Euclidean distance.
The red dot indicates the accuracy $\alpha$ of classification of the 1NN method based on the motif occurrence profile. The green dot indicates the accuracy $\beta$ of classification of the 1NN method based on the DTW. The blue dot indicates the accuracy $\gamma$ based on the Euclidean distance.
The ordinate represents the discriminant correctness rate based on the training set and the test set. The abscissa represents the time series length $L$.

In general, the DTW-based discriminant accuracy is the best and the motif profile method performs slightly worse, especially when the time series length is less than 200. The Euclidean distance method is the worst. Usually, the longer is the time series, the more information is extracted by the methods. However, the accuracy $\gamma$ of the classification method based on Euclidean distance decreases with the increase of the time series length. This is mainly because that the Euclidean distance calculation is simple. When the time series is longer, there is more noise, which is not conducive to depicting the similarity between time series. The Euclidean distance method has a relatively good effect when the time series length is less than 50. When the time series length is greater than 200, we find that $\alpha=1$ and $\beta=1$, indicating that the DTW-based method and the motif profile method provided in this paper are able to distinguish completely different time series. The Euclidean distance method is not good and it is the same as the random classification, from which the accuracy is $1/C$, where $C$ is the number of categories in the data set. The logistic map series has 5 categories, we have $1/C=1/5$. The discrete chaotic time series has 4 categories, we have $1/C=1/4$.

In order to analyze the accuracy of classification in the case of data loss, we perform the same analysis on the time series after data deletion. The length of the original time series is $L=512$. We randomly delete a proportion ($E$) of the data from each time series. We then classify the remaining data and calculate the classification accuracy rates. This process is repeated 10 times and the average classification accuracy rates $\alpha$, $\beta$ and $\gamma$ are obtained.
Fig.~\ref{Fig:TSMotif:Mimic:Classification} (B) and (D) show the relationship between the data deletion rate $E$ and the classification accuracy rates $\alpha$, $\beta$ and $\gamma$. Overall, $\alpha$, $\beta$ and $\gamma$ decrease with increasing $E$. We observe that the Euclidean distance method performs as the random classification, with $\gamma=1/5$ for the logistic maps and $\gamma=1/4$ for the chaotic time series. For the logistic maps, the motif profile method is more robust to data deletion than the DTW method. When the data deletion rate is close to 20\%, the motif-based classification accuracy $\alpha$ can still reach 100\%, while the DTW-based classification accuracy rate $\beta$ drops to about 80\%. In contrast, for the chaotic time series, the DTW method outperforms the motif profile method. The DTW classification method is very robust to data deletion and its accuracy rate $\beta$ is close to 100\% even when the data deletion rate $E$ is as high as 50\%. Not surprisingly, each method (DTW or motif profile) has its own advantages and disadvantages. Different methods usually have different performances when they are applied to different time series.

\section{Triadic time series motif analysis of the UCR Time Series Classification Archive}

To test the effectiveness of this method on similarity measures of time series, we use this method to classify real time series. The data source is from the UCR Time Series Classification Archive \citep{UCRArchive2018}.
The UCR Time Series Classification Archive contains 128 data sets, each of which is divided into a training set and a test set.
The dataset website also presents some results about classification accuracy of three methods. The first method uses the nearest neighbor method to classify based on the Euclidean distance. The accuracy rate is expressed by $\gamma$, where the highest rate is 100\% (Coffee) and the lowest correct rate is 5.77\%
(PigAirwayPressure).
The second uses the nearest neighbor method to classify based on the DTW method. The correct rate is represented by $\beta$, where the highest is 100\% (Trace, Two Patterns, Plane, Coffee) and the lowest correct rate is 10.58\% (PigAirwayPressure). The third method is based on the improvement of DTW. From previous research results, we found that the DTW method can describe the similarity of time series very well, which is more effective than the Euclidean distance, but the complexity of the DTW method is too high. We apply our method to the 128 data sets and compared the results with those obtained from the first and second methods.

\begin{figure}[htb]
\centering
\includegraphics[width=15cm]{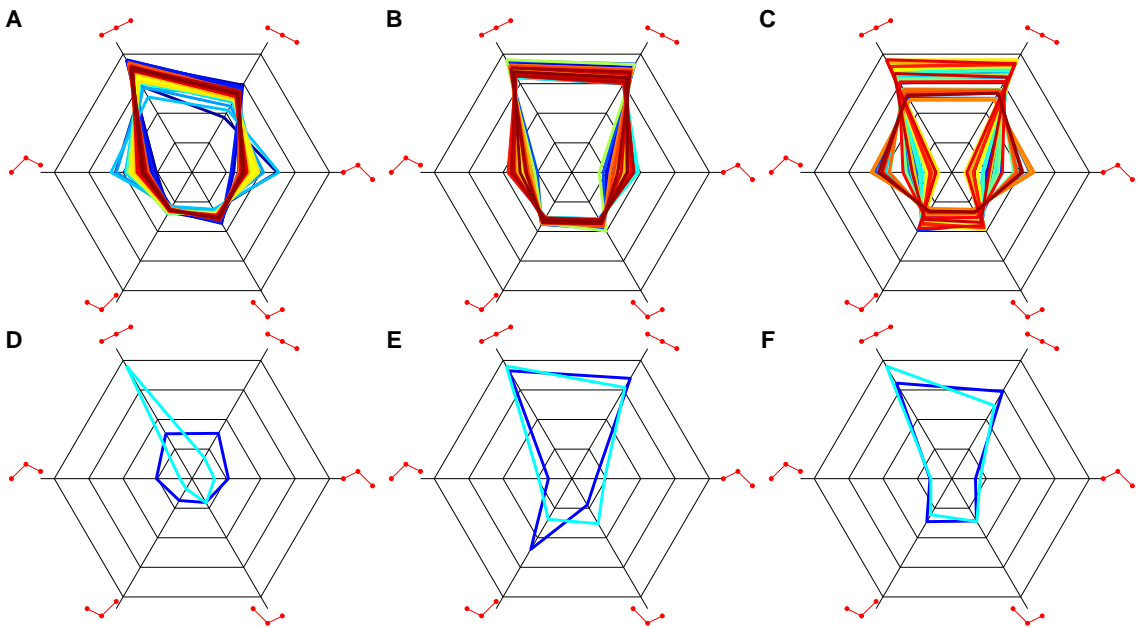}
\caption{Motif occurrence profiles $\mathbf{f}$ for different categories of time series in six data sets: PigAirwayPressure (A), PigCVP (B), Phoneme (C), Wafer (D), MoteStrain (E), and ToeSegmentation2 (F).
Each radar chart corresponds to a data set. Each solid line in the radar map represents the average motif occurrence profile of a class of time series in the data set.
}
\label{Fig:TSMotif:UCR:Radar}
\end{figure}

\begin{figure}[htb]
\centering
\includegraphics[width=8cm]{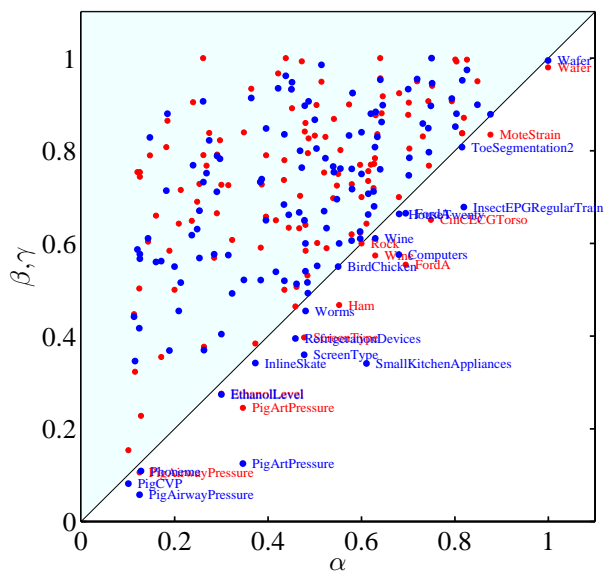}
\caption{Results of classification based on the occurrence frequency $f_{i}$ of motifs.
(A) Each data point in the figure corresponds to a data set. In the figure, the abscissa indicates the correct discriminant rate based on the motif distribution, $\alpha$, and the ordinate indicates the discriminant correctness rate, $\beta$, based on the DTW distance. (C) The abscissa in the figure indicates the correct discrimination rate based on the distribution of the motifs, $\alpha$, and the ordinate indicates the correctness rate based on the Euclidean distance, $\gamma$.   Euclidean distance, and DTW based.
}
\label{Fig:TSMotif:UCR:Acc}
\end{figure}

Fig.~\ref{Fig:TSMotif:UCR:Radar} shows the radar charts of the motif occurrence profiles averaged within different classes of time series in six representative data sets. Each radar chart corresponds to a data set. Each solid line in the radar map represents the average of the motif occurrence profile $\mathbf{f}$ of one category of time series in the data set. The time series belonging to the same class in the training set and the test set are included in the averaging process.
%
%
It can be seen that the six radar charts are very different, implying that the motif profile method can classify different data sets effectively. For each data set, the difference between the profile lines in the corresponding radar chart represents the difference between different time series. The classification will be more accurate if the difference is larger. The three radar charts on the top panel of Fig.~\ref{Fig:TSMotif:UCR:Radar} have many profile lines that are not sufficiently separated, which indicates that it would be hard to distinguish those categories. In contrast, each of the three radar charts on the bottom panel of Fig.~\ref{Fig:TSMotif:UCR:Radar} have only two profile lines that are well separated, which indicates that the two categories can be well distinguished. Indeed, the classification accuracy is low for the former data sets and high for the later data sets (see also Fig.~\ref{Fig:TSMotif:UCR:Acc}).


We use the triadic motif occurrence profile as the characteristic time series feature to classify the 128 data sets. The classification accuracy is shown in Fig. \ref{Fig:TSMotif:UCR:Acc}.
For 11 data sets, our method is better than the DTW method, since $\alpha \geq \beta$. The 11 data sets is shown in the lower right triangle of Fig.~\ref{Fig:TSMotif:UCR:Acc}.
Fig.~\ref{Fig:TSMotif:UCR:Acc} also compares the classification accuracy of the motif profile method and the Euclidean distance method. There are 18 data sets satisfying $\alpha \geq \gamma$, indicating that our method performs better than the Euclidean distance method for these data sets.
For instance, for the data set \emph{SmallKitchenAppliances}, our method is 50\% more accurate than the Euclidean distance method.
In general, the DTW method does a very good job in the measurement of time series similarity. Our method is superior to the DTW method for some data sets.

\section{Conclusions}

It is of great significance to be able to discover the characteristics of time series from a unique perspective through novel methods. Here, we studied the characteristics of time series through triadic time series motifs. We defined six different network motif. The simulation analysis finds that the distributions of the motif occurrence frequencies corresponding to logistic maps and chaotic time series (chaotic Henon map, chaotic Ikeda map, hyperchaotic generalized Henon map and hyperchaotic folded-tower map) all have their own characteristics. The motif occurrence profiles can quantify the time series characteristics in different dynamical systems and show comparative classification power as the DTW method.

We apply the motif analysis to the UCR data sets. The advantage of the Euclidean distance method is that the calculation is simple and fast. The DTW method performs best, but in some data sets, the performance is not as good as the motif profile method. Our method has better accuracy than the DTW method for 11 data sets.

The starting point of our method is completely different from the Euclidean distance method and the DTW method. This study is based on the complex networks, and mines the features in the time series. It is expected to be effectively improved in future research and provide a more effective method for measuring the similarity of time series. Indeed, there are many methods for extracting motifs from time series. Different motif extraction methods can describe different time series features. In order to improve the practicality of our method, we will develop different motif recognition methods to measure time series similarity.

\section*{Acknowledgements}

This work was supported by National Natural Science Foundation of China (11505063, 71532009, U1811462) and Fundamental Research Funds for the Central Universities (222201818006).


\end{document}